\documentclass[12pt,a4paper]{article}

\usepackage{latexsym}
\usepackage{setspace}
\usepackage{a4wide}
\usepackage{amsmath}
\usepackage{amsfonts}
\usepackage{amscd}
\usepackage{amssymb}
\usepackage{amsbsy}
\usepackage{mathrsfs}
\usepackage{yfonts}
\usepackage{bbm}
\usepackage[dvips]{epsfig,graphics}
\usepackage{epsfig}
\usepackage{graphicx}
\usepackage{psfrag}
\usepackage{upgreek}
\usepackage{undertilde}
\usepackage{color}
\usepackage{leftidx}
\usepackage{mathbbol}

\DeclareMathAlphabet{\mathpzc}{OT1}{pzc}{m}{it}

\begin{document}

{\color{red}{This document is the Accepted Manuscript version of a Published Work that appeared in final form in International Journal of Engineering Science, copyright Elsevier after peer review and technical editing by the publisher. To access the final edited and published work see \\ http://www.sciencedirect.com/science/article/pii/S0020722514001797}}

\title{Curvature dependent surface energy for free standing monolayer graphene: geometrical and material linearization with closed form solutions}

\author{D. Sfyris, G.I. Sfyris, C. Galiotis}

\maketitle

\begin{abstract}

Continuum modeling of a free-standing graphene monolayer, viewed as a two dimensional 2-lattice, requires specifications of the components of the shift vector that act as an auxiliary variable. The field equations are then the equations ruling the shift vector, together with momentum and moment of momentum equations. To introduce material linearity energy is assumed to have a quadratic dependence on the strain tensor, the curvature tensor, the shift vector, as well as to combinations of them. Hexagonal symmetry then reduces the overall number of independent material constants to nine. We present an analysis of simple loading histories such as axial, biaxial tension/compression and simple shear for a range of problems of increasing difficulty for the geometrically and materially linear case. We start with the problem of in-plane motions only. By prescribing the displacement, the components of the shift vector are evaluated. This way the field equations are satisfied trivially. Out-of-plane motions are treated as well; we assume in-plane tension/compression that leads to buckling/wrinkling and solve for the components of the shift vector as well as the function present in buckling's modeling. The assumptions of linearity adopted here simplifies the analysis and facilitates analytical results. 

\end{abstract}

\textbf{Keywords:} monolayer graphene; tension/compression; simple shear; geometrical linearities; material linearities; monoatomic 2-lattice.

\section{Introduction}

Graphene is a two dimensional sheet that constitutes the building unit of all graphitic forms of matter, such as graphite, carbon nanotubes and carbon fibers. Lee etal. (\cite{Leeetal2008}) use a nanoidentation experiment in an atomic force microscope to measure the elastic properties and intrinsic strength of graphene. Using second order elasticity they evaluate Young's modulus, the second order elastic constant as well as graphene's breaking strength. Their analysis models graphene as an isotropic body in one dimension, due to symmetry in the loading. 

Generalization of their approach to two dimensions is done by Cadelano et al. (\cite{Cadelanoetal2009}). These authors view graphene as an isotropic body and they utilize an energy cubic in strains (second order elasticity in words of Murnaghan and Rivlin \cite{Rivlin1963,Murnaghan1951}). Utilizing tight-binding atomistic simulations they calculate Young's modulus, Poisson ratio as well as higher order constants for graphene. While interesting and novel their approach is, it lacks the treatment of bending effects. It also models graphene as an isotropic body; dependence on the zig-zag and the armchair direction is not incorporated to the constitutive law through dependence on a structural tensor. Fifth order models for graphene are presented by Wei et al. (\cite{Weietal2009}). These authors utilize an energy that depends on strains of the fifth order. Using density functional theory for simple loading histories they evaluate higher order constants for graphene. Their approach does not include bending effects neither anisotropy; graphene is modeled as an isotropic body. 

To introduce anisotropy for a free-standing monolayer graphene as well as for incorporating bending effects we recently proposed a finite elasticity model for graphene (\cite{Sfyris-Galiotis2014}). Viewing graphene as a two dimensional 2-lattice, we obtain its arithmetic symmetries (\cite{Fadda-Zanzotto2000,Pitteri-Zanzotto2003}). Confined to weak transformation neighborhoods (\cite{Ericksen1979,Pitteri1985}) and invoking the Cauchy-Born rule (\cite{Ericksen2008}), we arrive to the classical symmetries continuum mechanics uses. We lay down the complete and irreducible representation (\cite{Zheng1994,Zheng1997}) for an energy depending on the Cauchy-Green deformation tensor, the curvature tensor as well as the shift vector. Cauchy-Green's surface tensor is a measure of in-plane motions, the curvature tensor measures out-of-plane motions, while dependence on the shift vector stems from viewing graphene as a 2-lattice. Dependence of the energy on the curvature tensor is motivated by the fundamental works of Murdoch and Cohen, Steigmann and Ogden (\cite{Murdoch-Cohen1979,Steigmann-Ogden1999}). We note that E and Ming (\cite{E-Ming2007}) report dependence on the energy on the shift vector for graphene as well. The need for introducing the shift vector as an independent variable in continuum modeling of graphene is also apparent in the approach of Zhu and Huang (\cite{Zhou-Huang2008}). Additionally, the corrugation vector that is introduced in the homogenization scheme of Davini (\cite{Davini2014}) is very close in spirit to the shift vector of our approach. 

In \cite{Sfyris-Galiotis2014} anisotropy is introduced throught a sixth-order strucural tensor which describes the zig-zag and armchair directions of graphene. This model predicts 13 independent material moduli, in contrast to the seemingly endless Taylor expansion models in terms of the strains adopted at third and fifth order elasticity \cite{Cadelanoetal2009,Weietal2009}. It is worth mentioning that bending effects are considered in the work of Wei et al. (\cite{Weietal2013}). These authors utilize an energy depending on one in-plane measure and two out-of-plane: bending rigidity and Gaussian bending stifness. These two quantities are work conjugate to the mean and the Gaussian curvature, respectively. Using density functional calcualtions for single wall carbon nanotubes, they evaluate bending rigidity and Gaussian bending stifness for a monolayer graphene. Their calculations are based on assuming infinitely long constant radius carbon nanotubes, so thay can relate energy per atom of the carbon nanotube to the energy per atom of the graphene sheet. 

Another interesting study incorporating bending effects is that of Lu-Huang (\cite{Lu-Huang2009}). Using von-Karman kinematical assumptions together with a measure of curvature they provide stress-strain curves using the virial theorem and molecular calculations. In-plane constants are calculated together with bending stiffness which is work-conjugate to curvature. Mixed atomistic-continuum methods are reported by Arroyo and Belytscko (\cite{Arroyo-Belytschko2002,Arroyo-Belytschko2004}) based on the earlier notion of the quasicontinuum (\cite{Tadmoretal1996,Tadmoretal1999}). Arroyo and Belytschko provide a finite continuum theory derived from interatomic potential; the material moduli are expressed in an explicit form in terms of the interatomic potential. They also provide a generalization of the Cauchy-Born rule appropriate for modeling surfaces. 

The present work is the linearized counterpart of our previous contributions (\cite{Sfyris-Galiotis2014,Sfyrisetal2014}). Linearization is understood at two levels: material linearity as well as geometrical linearity. Geometrical linearity means confinement to small deformations; mathematically this means that higher order terms of the displacement gradient are negligible. Material linearity means that energy is a quadratic function of the strain tensor, the curvature tensor, the shift vector, as well as to combinations of them. Anisotropy is introduced by requiring the tensors of material constants to be independent under rotations by 60$^0$: graphene's symmetry. This reduces the independent moduli to 9. 

We then examine what this framework gives for simple loading histories. Initially, we treat the case of in-plane deformations only. We thus disregard out-of-plane effects setting the curvature tensor equal to zero. In this case we need not take into account the equations of moment of momentum. Assuming the form of the displacement components that correspond to axial tension/compression, we solve for the components of the shift vector. It turns out that shift's vector components are homogeneous; they depend on the loading parameter as well as on the material constants. Same homogeneity of the shift vector components holds true for the case of biaxial tension/compression and for the simple shear case. Analogous procedure is done in the nonlinear counterpart of the present theory (\cite{Sfyrisetal2014}). Results there (\cite{Sfyrisetal2014}) are obtained using the same procedure, nevertheless they are much more complicated than the results of this study. This is due to the linearity assumptions that simplify the analysis here severely. This is apparent especially in the equations describing the shift vector. In the linearized problem they are algebraic equations of the first order, while for the nonlinear case they are algebraic equations of the fifth order. This order reduction simplifies the analysis and facilitates analytical results.

This difference in the algebraic nature of the equations ruling the shift vector permit closed form solutions for the buckling/wrinkling case as well, in contrast to the nonlinear case. By making a suitable assumption for the buckling mode (\cite{Punteletal2011}), we solve for the components of the shift vector. These expressions are substituted to one of the momentum equation. From this equation we obtain the form the function present in the buckling mode has. Then, this final expression is substituted to all the other field equations thereby rendering constraints that the material parameters, the loading constant and the constants of integration should satisfy so that all field equations are satisfied. 

The paper is structured as follows. Section 2 reminds the modeling of graphene as a 2-lattice, as well as the passage to the continuum theory. The field equations as well as the constitutive laws that introduce material linearity are given there. Section 3 deals with evaluating the number of independent constants for the constitutive law. Following standard approaches on the topic (see e.g. \cite{Nye1969}), we postulate invariance of the material tensors under rotations by 60$^0$: this is graphene's symmetry group. This reduces the number of independent moduli to 9. 

Section 5 deals with in-plane motions only. We disregard out-of-plane motions so the equations of moment of momentum are redundant, as is the curvature tensor. Making suitable assumptions for the displacement field describing axial, and biaxial tension/compression, we evaluate the components of the shift vector in order all field equations to be satisfied. Section 6 deals with buckling/wrinkling: we study in-plane deformations that ultimately lead to wrinkling/buckling. Evaluating the components of the curvature tensor that correspond to such a displacement, we search for the componets of the shift vector. When the latter are substituted to the momentum equations we obtain an equation for evaluating the function present in the buckling/wrinkling mode. We solve for this function and then make sure that all other field equations are satisfied. The paper ends up at Section 7 with some concluding remarks. 

As far as notation is concerned Greek indices range from 1 to 2. The common dot product is denoted by $\cdot$, tensor product by $\otimes$ while the cross product for the three dimensional space by $\times$. Summation of repeated indices is assumed throughout the paper. Initially, graphene is assumed to be a flat surface; namely a plane. 

\section{Graphene as a 2-lattice}
Following the  classification of 2-lattices by Fadda and Zanzotto (\cite{Fadda-Zanzotto2000}), we treat a monolayer graphene as a hexagonal monoatomic 2-lattice with unit cell of the form of Figure 1. 
\begin{figure}[!htb]
\centering
\includegraphics{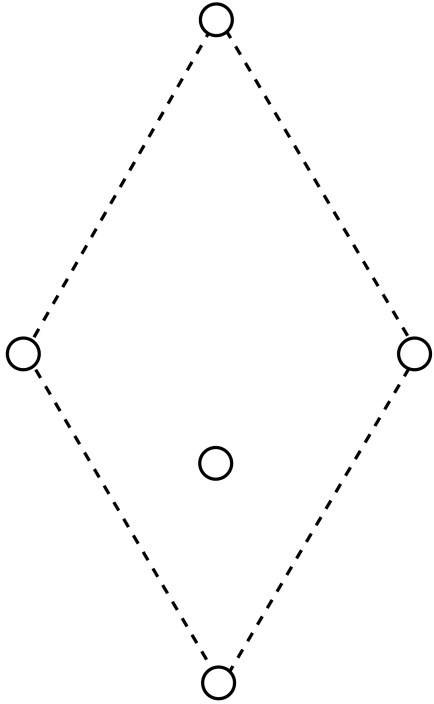}
\caption{The unit cell of a hexagonal 2-lattice (\cite{Fadda-Zanzotto2000}). }
\label{fig:digraph}
\end{figure}
The lattice and shift vectors are depicted in Figure 2
\begin{figure}[!htb]
\centering
\includegraphics{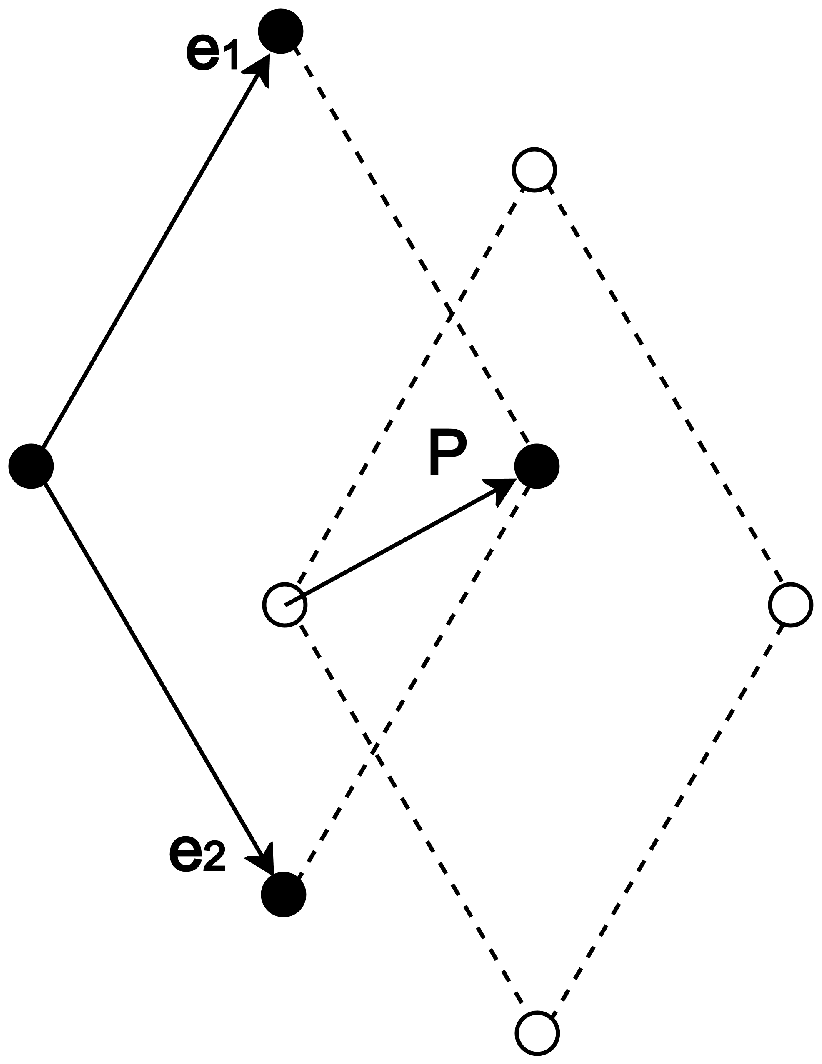}
\caption{The lattice and shift vectors of graphene. }
\label{fig:digraph}
\end{figure}
and defined as 
\begin{equation}
{\bf e}_1=(\sqrt{3} l, 0), \ \ {\bf e}_2=\left( \frac{\sqrt{3}}{2} l, \frac{3}{2} l \right), \ \ {\bf p}=\left( \frac{\sqrt{3}}{2} l, \frac{1}{2} l \right),
\end{equation}
$l$ being the lattice size, namely the interatomic distance at ease which is approximately 1, 42 Angstrom. The two simple hexagonal lattices are 
\begin{eqnarray}
&&L_1 (l)=\{ {\bf x} \in \mathcal R^2: {\bf x}=n^1 {\bf e}_1 + n^2 {\bf e}_2, \ \ (n^1, n^2) \in \mathcal Z^2 \}, \nonumber\\
&&L_2 (l)={\bf p}+L_1(l). 
\end{eqnarray}

The arithmetic symmetry group (\cite{Ericksen1979,Pitteri-Zanzotto2003}) of graphene is then described by the matrices 
\begin{equation}
\begin{pmatrix}
      -1 & -1 & -1 \\
      1 & 0 & 0 \\
      0 & 0 & 1
\end{pmatrix},
\begin{pmatrix}
      0 & 1 & 0 \\
      1 & 0 & 0 \\
      0 & 0 & 1
\end{pmatrix},
\begin{pmatrix}
      -1 & -1 & -1 \\
      0 & 1 & 0 \\
      0 & 0 & 1
\end{pmatrix},
\end{equation}
\begin{equation}
\begin{pmatrix}
      1 & 0 & 0 \\
      -1 & -1 & -1 \\
      0 & 0 & 1
\end{pmatrix},
\begin{pmatrix}
      1 & 0 & 0 \\
      0 & 1 & 0 \\
      0 & 0 & 1
\end{pmatrix},
\begin{pmatrix}
      0 & 1 & 0 \\
      -1 & -1 & -1 \\
      0 & 0 & 1
\end{pmatrix}.
\end{equation}
The eigenvalues of these matrices are $1, -1, e^{i \pi/3}, e^{-i \pi/3}$, so they describe the identity transformation, reflection transformation, and rotations by $60^0$, $-60^0$, respectively. 

At the continuum level, topologically, graphene is modeled as a two dimensional smooth surface embedded in a three dimensional Euclidean space. Position vectors on the reference configuration $\mathcal B_R$ of the referential surface are parametrized by two surface coordinates $\Theta^{\alpha}, \alpha=1, 2$ as (\cite{Ciarlet2005})
\begin{equation}
{\bf X}={\bf X}(\Theta^{\alpha}). 
\end{equation}  
After the deformation the surface occupies the current configuration $\mathcal B_C$, described by the position vector
\begin{equation}
{\bf x}={\bf x}(\Theta^{\alpha}). 
\end{equation} 
Covariant surface base vectors are then defined as 
\begin{equation}
{\bf A}_{\alpha}={\bf X}_{, \alpha }, \ \  {\boldsymbol \alpha}_{\alpha}={\bf x}_{, \alpha },
\end{equation}
for $\mathcal B_R$ and $\mathcal B_C$, respectively. Contravariant base vectors are given as
\begin{equation}
{\bf A}_{\alpha} \cdot {\bf A}^{\beta}=\delta_{\alpha}^{\beta}, \ \ {\boldsymbol \alpha}_{\alpha} \cdot {\boldsymbol \alpha}^{\beta}=\delta_{\alpha}^{\beta},
\end{equation}
$\delta_{\alpha}^{\beta}$ being the two dimensional Kronecker delta. 

The surface deformation gradient ${\bf F}_S$ reads
\begin{equation}
{\bf F}_S={\boldsymbol \alpha}_{\alpha} \otimes {\bf A}^{\alpha},
\end{equation}
while the surface right Cauchy-Green deformation tensor takes the form
\begin{equation}
{\bf C}_S={\bf F}_S^T \cdot {\bf F}_S.
\end{equation}
This tensor is related to the surface strain tensor by the formula 
\begin{equation}
{\bf e}=2 {\bf C}_S-{\bf I},
\end{equation}
$\bf I$ being the two dimensional unit tensor.

Geometrical linearity (small deformations) is introduced by defining the displacement vector ${\bf u}={\bf x}-{\bf X}$. Then, the deformation gradient reads ${\bf F}_S={\bf I}+\nabla_S {\bf u}$, with $\nabla_S()$ being the surface gradient defined as $\nabla_S()=\nabla()-{\bf n}({\bf n} \cdot \nabla())$, where $\bf n$ is the outward unit normal of the surface. Using this relation to eq. (10) together with eq. (11) one finally obtains for the strain tensor
\begin{equation}
e_{\alpha \beta}=\frac{1}{2}(u_{\alpha, \beta}+u_{\beta , \alpha}),
\end{equation}
when higher order terms, $u_{\alpha , \beta} u_{\alpha , \beta}$ are neglected due to the linear assumption. The geometrical linear case utilize the strain tensor $\bf e$ of eq. (12) which measure the in-plane deformations graphene suffers. Essentially, in this case the reference configuration $\mathcal B_R$ and current configuration $\mathcal B_C$ are very close to one another, so there is no need to distinguish between them.  
 
Out-of-plane deformations are described by the surface curvature tensor
\begin{equation}
{\bf b}=b_{\alpha \beta} {\boldsymbol \alpha}^{\alpha} \otimes {\boldsymbol \alpha}^{\beta}, 
\end{equation}
which is the second fundamental form of the surface. Taking into account bending effects for a monolayer graphene modeled as a surface, requires dependence of the energy on the curvature (\cite{Steigmann-Ogden1999,Murdoch-Cohen1979,Cohen-DeSilva1966}). Thus, for a monolayer graphene at the continuum level we assume an energy of the form (\cite{Sfyris-Galiotis2014,Sfyrisetal2014})
\begin{equation}
W=W({\bf e}, {\bf b}, {\bf p}). 
\end{equation}   
Dependence on the shift vector, $\bf p$, at the continuum level, results from the fact that at the crystalline level graphene is a 2-lattice. Now, we confine ourselves to weak transformation neighborhoods (\cite{Pitteri-Zanzotto2003}) and assume validity of the Cauchy-Born rule (\cite{Ericksen2008}). With these assumptions enforced we may utilize the classical symmetries employed by continuum mechanics.

Material linearity is introduced by quadratic dependence of the energy
\begin{eqnarray}
W({\bf e}, {\bf b}, {\bf p})&&=\frac{1}{2} C^1_{ijkl} e_{ij} e_{kl} +\frac{1}{2} C^2_{ij} p_i p_j +\frac{1}{2} C^3_{ijk} e_{ij} p_k \nonumber\\
&&+\frac{1}{2} C^4_{ijkl} b_{ij} b_{kl} +\frac{1}{2} C^5_{ijkl} e_{ij} b_{kl} +\frac{1}{2} C^6_{ijk} b_{ij} p_k. 
\end{eqnarray} 
Tensors ${\bf C}^1, {\bf C}^4, {\bf C}^5$ are fourth order tensors, ${\bf C}^3, {\bf C}^6$ are third order tensors, while ${\bf C}^2$ is a second order tensor: all these tensors are tensors of material parameters. The components of ${\bf C}^1$ describe pure in-plane moduli, those of ${\bf C}^4$ pure out-of-plane moduli, while those of ${\bf C}^5$ mixed in-plane with out-of-plane moduli. Components of ${\bf C}^3, {\bf C}^6$ describe the effect of strain and curvature, respectively, on the shift vector. Finally, ${\bf C}^2$ gives the material modulus related with the shift vector's motions, solely. 

The field equations for such a problem are the momentum equation, the moment of momentum equation as well as the equations ruling the shift vector. For the momentum equation we have (\cite{Chhapadiaetal2011,Sfyris-Galiotis2014}) when body forces and inertia are absent
\begin{equation}
\boldsymbol \sigma^{\textrm{bulk}} \cdot {\bf n}+\nabla_S {\boldsymbol \sigma}=0,
\end{equation}
where $\boldsymbol \sigma$ is Cauchy's stress tensor for the surface, while $\boldsymbol \sigma^{\textrm{bulk}}$ is the stress tensor of the bulk material. Here the sheet of graphene is assumed to be free-standing, so $\boldsymbol \sigma^{\textrm{bulk}}$ is set equal to zero. Since we confine ourselves to small deformations we need not distinguish between different stress measures for the surface stress measures. The moment of momentum equation in the absence of body couples, inertia and bulk material reads
\begin{equation}
\textrm{div} {\bf m}-\nabla ({\boldsymbol \sigma} \times {\bf u})=0, 
\end{equation}
where $\bf m$ is the surface couple stress tensor. For the shift vector the field equation reads (\cite{Pitteri-Zanzotto2003,E-Ming2007})
\begin{equation}
\frac{\partial W}{\partial {\bf p}}={\bf 0}.
\end{equation}
Form the physical point of view, the momentum equation is the force balance for the surface, while the moment of momentum renders the couple balance for the surface. The shift vector adjusts according to eq. (18) in order equilibrium to be reached (\cite{Pitteri-Zanzotto2003}). 

\section{Constitutive relations}

For obtaining the exact form of the constitutive relations we need to evaluate the independent components of the tensors ${\bf C}^1,..., {\bf C}^6$. Symmetries of graphene (see eqs. (3, 4)) dictate that they should be invariant under rotations by 60$^0$. Certainly, the arithmetic symmetries  of eqs. (3, 4) are for the atomistic point of view. Passage to the continuum requires confinement to weak transformation neighborhoods as well as enforcement of the Cauchy-Born rule (see \cite{Sfyris-Galiotis2014} and references therein). 

For evaluating the independent constants, we start with the following systems for tensors of fourth, third and second order, respectively (\cite{Nye1969})
\begin{eqnarray}
\tilde{C}_{ijkl}&&=a_{ip} a_{jq} a_{kr} a_{ls} C_{pqrs}, \\
\tilde{C}_{ijk}&&=a_{ip} a_{jq} a_{kr} C_{pqr}, \\
\tilde{C}_{ij}&&=a_{ip} a_{jq}  C_{pq},
\end{eqnarray}
where the tensor $\bf a$ describe rotation by 60$^0$ and has the following matrix form 
\begin{equation}
[a_{ij}]=\begin{pmatrix}
      \frac{1}{2} & \frac{\sqrt{3}}{2}  \\
      -\frac{\sqrt{3}}{2} & \frac{1}{2} \\
\end{pmatrix}.
\end{equation}
Using eq. (22) on eqs. (19-21) one obtains systems for the components of the material moduli. Then, setting (\cite{Nye1969})
\begin{equation}
\tilde{C}_{ijkl}=C_{ijkl}, \ \  \tilde{C}_{ijk}=C_{ijk}, \ \ \tilde{C}_{ij}=C_{ij},
\end{equation} 
invariance of the material moduli under rotations by 60$^0$ is enforced. 

For the components of the fourth order tensors one finally obtains two independent moduli (\cite{Guinovartetal2001}). For the third order constants the independent moduli is one, as has been evaluated by Nye (\cite{Nye1969}, p. 124, Table 8) for piezoelectric problems. For the second order tensor one component is independent as one can evaluate. All in all, the constitutive expression for the surface stress then read
\begin{eqnarray}
\sigma_{11}&&=c_1 e_{11}+c_2 e_{22} + c_3 b_{11} + c_4 b_{22} -c_5 p_2, \\
\sigma_{22}&&=c_2 e_{11} + c_1 e_{22}+c_4 b_{11}+c_3 b_{22}+c_5 p_2, \\
\sigma_{12}&&=\frac{c_1-c_2}{2} e_{12}+\frac{c_3-c_4}{2} b_{12}-2 c_5 p_1,
\end{eqnarray}
steming from the expression 
\begin{equation}
{\boldsymbol \sigma}=\frac{\partial W}{ \partial {\bf e}}=C^1_{ijkl} e_{kl}+C^3_{ijk} p_k+C^5_{ijkl} b_{kl}.
\end{equation}
The constants $c_1, c_2$ are the independent moduli of the tensor $C^1$, $c_3, c_4$ is related with $C^3$ while $c_5$ stems from $C^5$.

For the surface  couple stress the constitutive law reads
\begin{equation}
{\bf m}=\frac{\partial W}{\partial {\bf b}}=C^4_{ijkl} b_{kl} +C^5_{ijkl} e_{kl} +C^6_{ijk} p_k.
\end{equation}
So, we obtain
\begin{eqnarray}
m_{11}&&=c_6 b_{11}+c_7 b_{22}+c_3 e_{11}+c_4 e_{22}-c_8 p_2, \\
m_{22}&&=c_7 b_{11}+c_6 b_{22}+c_4 e_{11}+c-3 e_{22}+c_8 p_2, \\
m_{12}&&=\frac{c_6-c_7}{2} b_{12}+\frac{c_3-c_4}{2} e_{12}-2c_8 p_1.
\end{eqnarray}
The material parameters $c_6, c_7$ are related to $C^4$, while $c_8$ is related to $C^6$.

For the components related with the shift vector we have
\begin{equation}
\frac{\partial W}{\partial p_i}=C^2_{ij} p_j+C^3_{ijk} e_{jk} +C^6_{ijk} b_{jk}.
\end{equation}
So, we finally take
\begin{eqnarray}
\frac{\partial W}{\partial p_1}=&&c_9 p_1-2 c_5 e_{12}-2c_8 b_{12}, \\
\frac{\partial W}{\partial p_2}=&&c_9 p_2 -c_5 e_{11}+c_5 e_{22}-c_8 b_{11}+c_8 b_{22},
\end{eqnarray}
where $c_9$ is the independent moduli related to $C^2$.

\section{Field equations}

Using eqs. (24-26) to eq. (16) we obtain for the momentum equation
\begin{eqnarray}
&&c_1 u_{1,11}+c_2 u_{2,21}+c_3 b_{11,1}+c_4 b_{22,1}-c_5 p_{2,1}+\frac{c_1-c_2}{4} (u_{1,22}+u_{2,12}) \nonumber\\
&&\ \ \ \ \ \ \ \ \ \ \ \ \ \ \ \ \ \ \ \ \ \ \ \ \ \ \ \ \ \ +\frac{c_3-c_4}{2} b_{12,2}-2 c_5 p_{1,2}=0, \\
&&\frac{c_1-c_2}{4} (u_{1,21}+u_{2,11})+\frac{c_3-c_4}{2} b_{12,1} -2 c_5 p_{1,1} +c-2 u_{1,12}+c_1 u_{2,22} \nonumber\\
&&\ \ \ \ \ \ \ \ \ \ \ \ \ \ \ \ \ \ \ \ \ \ \ \ \ \ \ \ \ \ +c_4 b_{11,2}+c_3 b_{22,2}+c_5 p_{2,2}=0. 
\end{eqnarray}
After using of eqs. (33, 34) to eq. (18) the equations ruling the shift vector render
\begin{eqnarray}
&& c_9 p_1 -c_5 (u_{1,2} +u_{2,1}) -2 c_8 b_{12}=0, \\
&& c_9 p_2 -c_5 u_{1,1} +c_5 u_{2,2} -c_8 b_{11}+c_8 b_{22}=0. 
\end{eqnarray}

For the moment of momentum equations we use eqs. (24-26), (29-31) to eq. (17) and obtain
\begin{eqnarray}
&& \left[ c_6 b_{11}+c_7 b_{22}+c_3 u_{1,1}+c_4 u_{2,2}-c_8 p_2 \right]_{,1} +\left[ \frac{c_1-c_2}{2}b_{12}+\frac{c_3-c_4}{4} (u_{1,2}+u_{2,1}-2 c_8 p_1) \right]_{,2} \nonumber\\
&& -\left( u_1 [\frac{c_1-c_2}{4}(u_{1,2}+u_{2,1})+\frac{c_3-c_4}{2} b_{12}-2c_5 p_1]   \right)_{,1} \nonumber\\
&& -\left( u_1 [\frac{c_1-c_2}{4}(u_{1,2}+u_{2,1})+\frac{c_3-c_4}{2} b_{12}-2c_5 p_1]   \right)_{,2} \nonumber\\
&& +\left( u_2 [c_1 u_{1,1}+c_2 u_{2,2}+c_3 b_{11}+c_4 b_{22}-c_5 p_2]    \right)_{,1} \\
&& +\left( u_2 [c_1 u_{1,1}+c_2 u_{2,2}+c_3 b_{11}+c_4 b_{22}-c_5 p_2]    \right)_{,2}=0 \nonumber
\end{eqnarray}
and 
\begin{eqnarray}
&& \left[ \frac{c_6-c_7}{2} b_{12}+\frac{c_3-c_4}{4} (u_{1,2}+u_{2,1})-2 c_8 p_1  \right]_{,1} + \left[ c_7 b_{11}+c_6 b_{22}+c_4 u_{1,1}+c_3 u_{2,2}+c_8 p_2  \right]_{,2} \nonumber\\
&& -\left( u_1 [ c_2 u_{1,1} +c_1 u_{2,2}+c_4 b_{11}+c_3 b_{22}+c_5 p_2  ]  \right)_{,1} \nonumber\\
&& -\left( u_1 [ c_2 u_{1,1} +c_1 u_{2,2}+c_4 b_{11}+c_3 b_{22}+c_5 p_2  ]  \right)_{,2} \nonumber\\
&& +\left( u_2 [ \frac{c_1-c_2}{4} (u_{1,2}+u_{2,1})+\frac{c_3-c_4}{2} b_{12}-2c_5 p_1   ]    \right)_{,1} \\
&& +\left( u_2 [ \frac{c_1-c_2}{4} (u_{1,2}+u_{2,1})+\frac{c_3-c_4}{2} b_{12}-2c_5 p_1   ]    \right)_{,2}=0. \nonumber
\end{eqnarray}
Eqs. (35-40) are the counterpart of eqs. (16-18) written in terms of the kinematical measures: ${\bf u}, {\bf p}, {\bf b}$, which are the unknown functions.

\section{In-plane motions only}

When in-plane motions are considered only, the curvature tensor should be set equal to zero. Also, the moment of momentum equation need not be taken into account. The field equations for this case therefore, read
\begin{eqnarray}
&& c_1 u_{1,11} +c_2 u_{2,21}-c_5 p_{2,1} +\frac{c_1-c_4}{4} [u_{1,22}+u_{2,12}]-2c_5 p_{1,2}=0, \\
&& \frac{c_1-c_2}{4} [u_{1,21}+u_{2,11}]-2c_5 p_{1,1}+c_2 u_{1,12}+c_1 u_{2,22}+c_5 p_{2,2} =0, \\
&& c_9 p_1 -c_5 (u_{1,2}+u_{2,1})=0, \\
&& c_9 p_2 -c_5 u_{1,1}+c_5 u_{2,2} =0.
\end{eqnarray}
The first two are the momentum equations while the rest are the equations ruling the auxiliary variables. 

\subsection{Axial tension/compression}

For modeling axial tension/compression we assume for the displacement field
\begin{equation}
u_1=\epsilon \Theta^1, \ \  u_2 =  \Theta^2.
\end{equation}
This field of displacement models axial tension/compression in the $\Theta^1$ direction. When the loading constant $\epsilon$ is greater than zero, then we speak about tension, while when it is negative we speak about compression. The necessary derivatives for this case read
\begin{equation}
u_{1,1}=\epsilon, \ \ u_{1,2}=u_{2,1}=0, \ \ u_{2,2}=1. 
\end{equation}
The equations of the shift vector render
\begin{eqnarray}
&& c_9 p_1 =0 \rightarrow p_1=0, \\
&& c_9 p_2-c_5 \epsilon +c_5 =0 \rightarrow p_2 =\frac{c_5(\epsilon-1)}{c_9}.
\end{eqnarray}
So, the outcome for the one dimensional tension/compression leads to a homogeneous solution of the shift vector. With eqs. (47, 48) the momentum equations (eqs. (41, 42)) are satisfied trivially as one can infer by direct substitution. Thus, the pair $(p_1, p_2)=\left( 0, \frac{c_5(\epsilon-1)}{c_9} \right)$ qualifies as a solution for the problem at hand when loading is of the form of eq. (45).

\subsection{Biaxial tension/compression}

For modeling tension/compression in both directions we set for the displacement field
\begin{equation}
u_1 =\epsilon_1 \Theta^1, \ \ u_2 =\epsilon_2 \Theta^2.
\end{equation}
For the necessary derivatives we evaluate
\begin{equation}
u_{1,1} =\epsilon_1, \ \ u_{1,2}=0=u_{2,1}, \ \ u_{2,2}=\epsilon_2.
\end{equation}
The equations ruling the shift vector take then the form
\begin{eqnarray}
&& c_9 p_1 =0 \rightarrow p_1 =0, \\
&& c_9 p_2 -c_5 \epsilon_1 +c_5 \epsilon_2=0 \rightarrow p_2 =\frac{c_5 (\epsilon_2-\epsilon_1)}{c_9}.
\end{eqnarray}
Therefore, for the biaxial loading as well we obtain homogeneous solutions for the expressions of the components of the shift vector. Therefore, the momentum equations are satisfied trivially. Collectively, the pair $(p_1, p_2)=\left( 0,\frac{c_5 (\epsilon_2-\epsilon_1)}{c_9} \right)$ qualifies as a solution when loading is given by eq. (49).

\subsection{Simple shear}

Simple shear is described by a displacement field given by 
\begin{equation}
u_1 =\Theta^1 +\epsilon \Theta^2, \ \ u_2 =\Theta^2.
\end{equation}
For the derivatives we evaluate
\begin{equation}
u_{1,1}=1, \ \ u_{1,2}=\epsilon, \ \ u_{2,1}=0, \ \ u_{2,2}=1.
\end{equation}
The equations ruling the shift vector are then 
\begin{eqnarray}
&& c_9 p_1 -c_5 \epsilon =0 \rightarrow p_1 =\frac{c_5 \epsilon}{c_9} \\
&& c_9 p_2 -c_5+c_5=0 \rightarrow p_2 =0. 
\end{eqnarray}
The momentum equations are then satisfied trivially, since the solution in terms of the shift vector is homogeneous. All in all, the pair $(p_1, p_2)=\left( \frac{c_5 \epsilon}{c_9}, 0 \right)$ qualifies as a solution when simple shear is given by eq. (53). It is interesting to note that simple shear in the other direction will lead to the same result in terms of the components of the shift vector. 

\section{Out-of-plane motions}

\subsection{Introducing wrinkling/buckling}
In order to model wrinkling/buckling we need to assume that the out of plane displacement is given by the following expression (\cite{Timoshenko,Punteletal2011})
\begin{equation}
u_3=u_3(\Theta^1, \Theta^2)=\textrm{cos} \left( \frac{n \pi \Theta^1}{2 L_1}   \right) f(\Theta^2), 
\end{equation}
$n$ being the number of sinusoidal wave in the $\Theta^1$ direction and $f$ is an arbitrary function (see Figure 7
\begin{figure}[!htb]
\centering
\includegraphics{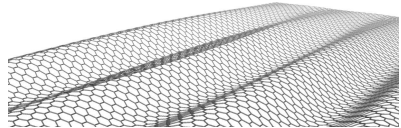}
\caption{Wrinkling/buckling described by eq. (57) (figure taken from \cite{Androulidakisetal2014}).}
\label{fig:digraph}
\end{figure}
for a schematic guide for this kind of deformation). The parametric form of a surface having the above expression as displacement is 
\begin{equation}
{\bf u}(\Theta^1, \Theta^2)=\left( \Theta^1, \Theta^2, \textrm{cos} \left( \frac{n \pi \Theta^1}{2 L_1}  \right) f(\Theta^2) \right).
\end{equation}
For our framework, $\bf b$ is the second fundamental form of the surface, so we evaluate for its components
\begin{eqnarray}
&& b_{11}={\bf u}_{,11} \cdot {\bf n} \\
&& b_{12}=b_{21}={\bf u}_{,12} \cdot {\bf n} \\
&& b_{22}={\bf u}_{,22} \cdot {\bf n}.
\end{eqnarray}
The outward unit normal, $\bf n$, of the surface is defined by 
\begin{equation}
{\bf n}=\frac{{\bf x}_{,1} \times  {\bf x}_{,2}  }{|{\bf x}_{,1} \times  {\bf x}_{,2}| }.
\end{equation}
These measures of the surface are important since they participate to the field equations (35-40) when out-of-plane motions are taken into acoount.

\subsection{Tension/Compression}

Axial tension/compression resulting in wrinkling/buckling is described by the parametric form of the surface 
\begin{equation}
{\bf u}(\Theta^1, \Theta^2)= \left( \epsilon \Theta^1, \Theta^2, \textrm{cos} \left( \frac{n \pi \Theta^1}{2 L_1} \right) f(\Theta^2) \right).
\end{equation}
Such an assumption means that tension/compression in the in-plane results in wrinkling/buckling, i.e. out of plane motion. The phenomenon is not assumed to be dynamic in order to have tension/compression initially that finally leads to wrinkling/buckling. The method is semi-inverse: we assume the form that the solution has in the final form. Tension will finally produce wrinkling on the material, while compression will lead to buckling. Certainly, one expects different behaviour in these two kind of loadings. Such a hardening response cannot be captured by the model in its present form; generalizations should be made which are outside the scope of this work. 

For the above given surface the outward unit normal has components
\begin{equation}
{\bf n}= \left( -\frac{n \pi}{2 L_1} \textrm{sin}   \left( \frac{n \pi \Theta^1}{2 L_1} \right) f(\Theta^2), -\epsilon \textrm{cos}   \left( \frac{n \pi \Theta^1}{2 L_1} \right) f'(\Theta^2), \varepsilon \right),
\end{equation}
when for its Euclidean length we assume it is unity: 
\begin{equation}
||{\bf n}||=\sqrt{\left[ -\frac{n \pi}{2 L_1} \textrm{sin}   \left( \frac{n \pi \Theta^1}{2 L_1} \right) f(\Theta^2) \right]^2+\left[ -\epsilon \textrm{cos}   \left( \frac{n \pi \Theta^1}{2 L_1} \right) f'(\Theta^2) \right]^2+\epsilon^2}=1.
\end{equation}
For the components of the second fundamental form we then obtain
\begin{eqnarray}
&& b_{11}=-\epsilon \frac{n^2 \pi^2}{4 L_1^2} \textrm{cos}   \left( \frac{n \pi \Theta^1}{2 L_1} \right) f(\Theta^2), \\
&& b_{12}=b_{21}=-\epsilon \frac{n \pi}{2 L_1} \textrm{sin}   \left( \frac{n \pi \Theta^1}{2 L_1} \right) f'(\Theta^2), \\
&& b_{22}=\epsilon \textrm{cos}   \left( \frac{n \pi \Theta^1}{2 L_1} \right) f''(\Theta^2).
\end{eqnarray}

Under these assumptions the equations ruling the shift vector, eqs. (37, 38), can be solved as
\begin{eqnarray}
&& p_1 =-2 \frac{c_8}{ c_9} \epsilon \frac{n \pi}{2 L_1^2} \textrm{cos}  \left( \frac{n \pi \Theta^1}{2 L_1} \right) f(\Theta^2), \\
&& p_2 =\frac{1}{c_9} \left( c_5 -c_5 \epsilon -c_8 \left[ \epsilon  \frac{n^2 \pi^2}{4 L_1^2} \textrm{cos}   \left( \frac{n \pi \Theta^1}{2 L_1} \right) f(\Theta^2)+ \epsilon \textrm{cos}   \left( \frac{n \pi \Theta^1}{2 L_1} \right) f''(\Theta^2) \right]    \right).
\end{eqnarray}
With these expressions for the components of the shift vector the second of the equations of momentum, eq. (36), render one differential equation for the function $f$  after elimination of the term with the cosinus, in the form
\begin{equation}
A f''(\Theta^2)+B f'(\Theta^2)=0,
\end{equation}
where $A=-[\frac{c_3-c_4}{2} - 4 \frac{c_5 c_8}{c_9}] \epsilon \frac{n^2 \pi^2}{4 L_1^2}-[c_4+\frac{c_5 c_8}{c_9}] \epsilon \frac{n^2 \pi^2 }{4 L_1^2}$ and $B=[c_3-\frac{c_5 c_8 }{c_9 }] \epsilon$. Solving eq. (71) we obtain 
\begin{equation}
f(\Theta^2)=-\frac{A}{B} e^{-\frac{A}{B} \Theta^2}h_1 +h_2,
\end{equation}
where $h_1, h_2$ are constants of integration. When this expression is substituted to the first of the momentum equation, eq. (35), one finally obtains
\begin{equation}
-A \epsilon \frac{n^3 \pi^3}{8 L_1^3} \left( \frac{A}{B} e^{-\frac{A}{B} \Theta^2}h_1 +h_2  \right) +(C+B) \epsilon \frac{n \pi}{2 L_1} \frac{A^3}{B^3}  e^{-\frac{A}{B} \Theta^2}h_1=0,
\end{equation}
where $C=\frac{c_3-c_4}{2}-\frac{4 c_5 c_8}{c_9}$. The latter equation should be viewed as a constraint on the material parameters, through the quantities $A, B, C$, the loading constant, $\epsilon$, and the constants of integration, $h_1, h_2$ in order to fulfill the second of the momentum equation. To this constraint two additional constraint equations should be added; these stem from the moment of momentum equations (eqs. (39, 40)) by substituting eqs. (63, 69, 70, 72). This would result, as for eq. (73), to two equations that the loading parameter, the material constants and the integration constants should satisfy in order the displacement field of eq. (63) to be a solution for the problem at hand. We refrain from writing down these two additional constraints resulting from the moment of momentum equations, but we mention that they can be obtained by direct substitution of eqs. (63, 69, 70, 72) to eqs. (39, 40).

\section{Conclusions}

This work constitutes an extension of \cite{Sfyris-Galiotis2014} in the direction of giving some closed form solutions for a free standing monolayer graphene. The approach is valid for the geometrical and material linear framework at the level of the continuum.

We start by presenting the framework of \cite{Sfyris-Galiotis2014} suitable for the geometrically and materially linear regime. For the case of in plane motions we examine one dimensional tension/compression along both directions of the surface as well as the case of biaxial tension/compression and simple shear. The outcome cosnists of homogeneous solutions for the components of the shift vector that depend on the material parameters and the loading constant. For modeling out of plane motions we describe how wrinkling/buckling can be introduced into the framework. We evaluate explicitly the components of the shift vector as well as those of the curvature tensor so that all field equations are satisfied. 

As for future directions, we consider that investigation of thin graphene sheets on subtrates constitutes a highly challenging theoretical and experimental problem. The linearized equations presented here together with the incorporation of substrate effects to the model, will make the present approach more relevant to actual experimental set-ups such as [1]. \\

\section{Acknowledgements}
This research has been co-financed by the European Union (European Social Fund - ESF) and Greek national funds through the Operational Program "Education and Lifelong Learning" of the National Strategic Reference Framework (NSRF) - Research Funding Program: ERC-10 "Deformation, Yield and Failure of Graphene and Graphene-based Nanocomposites". The financial support of the European Research Council through the projects ERC AdG 2013 (‘‘Tailor Graphene’’) is greatfully acknowledged.




\vspace{0.1cm}

D. Sfyris\\
FORTH-ICE/HT, Patras, Greece\\
dsfyris@iceht.forth.gr\\

\vspace{0.1cm}

G.I. Sfyris \\
LMS, Paris, France\\

\vspace{0.1cm}

C. Galiotis \\
FORTH-ICE/HT, and Department of Chemical Engineering, Patras, Greece\\

\end{document}